\documentstyle[aps,pre]{revtex} 
\preprint{}
\begin{document}
\newcommand{\simgeq}{\raisebox{-0.6 ex}{$\stackrel{>}{\sim}$}}
\newcommand{\simleq}{\raisebox{-0.6 ex}{$\stackrel{<}{\sim}$}}
\title{Estimation of light transport parameters in biological media using 
coherent backscattering}
\author{S. Anantha Ramakrishna\footnote{Current address : Raman Research Institute, Bangalore - 560 080, India; e-mail : sar@rri.ernet.in}  
and K. Divakara Rao\footnote{e-mail : kdivakar@cat.ernet.in}}
\address{Laser Programme, Centre for  Advanced Technology, Indore 
- 452 013, India}
\maketitle
\begin{abstract}
The suitability of using the angular peak shape of the coherent
backscattered light for estimating the light transport
parameters of biological media has been investigated. Milk and methylene
blue doped milk were used as tissue phantoms for the
measurements carried out with a He-Ne laser (632.8nm). Results
indicate that while the technique accurately estimates the
transport length, it can determine the absorption coefficient
only when the absorption is moderately high ($\alpha > 1 cm^{-1}$) for
the long transport lengths typical of tissues. Further, the possibility of 
determining the anisotropy factor by estimating the single scattering 
contribution to the diffuse background is examined. \\


\end{abstract}
\newpage

\section{Introduction}
The use of light and lasers in medicine have increased manifold
in the past decade. Enormous efforts have been devoted to the 
development of  new diagnostic techniques such as NIR imaging 
\cite{tuchin} and flourescence spectroscopy of tissues
\cite{kortum}, as well as therapeutic uses such as photothermal 
coagulation and Photo-dynamic therapy \cite{grossweiner}. 
Tissues are highly scattering media
and an accurate knowledge of light transport parameters of the tissue 
is indispensible to describe the propagation of light in these media.
Several techniques have been developed to measure the light transport 
parameters in tissues \cite{wilson}. Apart from these, the use of the Coherent 
Backscattered peak from tissues to estimate the transport 
parameters has also been suggested \cite{yoo1,yoo2,yoon}.\\

The phenomenon of Coherent Backscattering (CBS) of light by
random media has attracted considerable attention since 1985
\cite{albada,wolf1}, when connection
was first made between elastic multiple scattering and Anderson
localization. This phenomenon also termed weak localization,
shows up as a sharp peak in the backscattered direction within a
narrow cone of angles. The angular width of the CBS peak is
mainly determined by the transport length of light in the medium
and the shape of the peak is slightly modified by the
absorption. This technique was used to characterize
the transport lengths of samples used in experiments on strong
localization of light \cite{wiersma1}. Etemad {\it et. al.,}\cite{etemad1} were
the first to study the effects of absorption in random media on
the CBS peak.\\

There have been some attempts to study the CBS peak from
biological tissues and obtain the transport properties 
\cite{yoo1,yoo2,yoon}.
These studies, however, have been inaccurate because of a
misinterpretation \cite{error} in applying Akkerman's
expression \cite{akkermans} derived for conservative,
semi-infinite media to absorptive media.  
Hence a systematic study
of the CBS peaks from biological tissues has not yet been made. Since this
is a non-contact method which can be used for online in-vivo
measurements also, it is important that a systematic evaluation
of this technique for the measurement of transport parameters be
carried out.  Eddowes {\it et. al.,}\cite{eddowes} have suggested that accurate
absorption information may be obtained by using Monte-Carlo
simulations to fit experimental measurements and have developed
Monte-Carlo based routines to identify the optical coefficients
for a given CBS angular profile. However the measurement of the
absorption parameter is dependent on both the transport length
and the limitations of the experimental setup, as will be explained,
and accurate absorption information can be obtained in
only limited ranges of the absorption and transport
coefficients.  In reports that have appeared so far, researchers
have concentrated on obtaining the transport and inelastic
lengths of tissues from the shape of the CBS angular profile.
The possibility of using the intensity information for
estimating the anisotropy factor($g$) of the scatterers has not
been examined.\\

In this paper, we present the results of our investigations on
the suitability of using CBS for measuring transport
parameters of biological media.  
Milk of different concentrations have been used as
tissue phantom.  It is known that the light transport properties
of milk closely resemble those of tissues and milk has been used
to model light transport in brain tissue \cite{greenfield}, to
study optical imaging \cite{morgan}, photon-density
waves \cite{knuttel}, and for investigating the propagation of
short laser pulses through a scattering medium \cite{tereshchenko}.
To investigate the effects of absorption on CBS, an
absorbing dye with known extinction coefficient was added to
milk and the CBS from such media were studied. We also examine
the possibility of determining the anisotropy factor ($g$) from
the the single scattering contribution to the diffuse
background.\\

\section{Coherent Backscattering of light} 

Coherent Backscattering (CBS) of light occurs in all disordered
media and is the only major surviving interference effect.  When
a beam of light is incident on a random medium, there exist
partial waves traversing every possible path in the medium.  The
CBS effect arises from the constructive interference of any
partial wave with its time reversed counterpart in the medium.
In exactly the backscattered direction, both these two waves
have the same phase and constructive interference results. Away
from the backscattered direction, the counterpropagating paths
develop a phase difference depending on the relative positions
of the first and last scattering events in the medium. For the
ensemble of all possible light paths, these phases will
randomize and the reflection is enhanced within a narrow cone in
the backward direction with an angular width of the order of
$\lambda / {l_t} $ where $\lambda$ is the wavelength of light
and $l_t$ is the transport length in the medium. This peak shows
up only after the ensemble averaging over the large scale sample
specific fluctuations (speckle) that originate from the random
medium \cite{etemad2}.\\

The CBS intensity can be described in terms of three contributions. 
The total normalized angular intensity is described \cite{wiersma2} 
by
\begin{equation}
I(\theta, L) = \frac{\gamma_c(\theta, L) + \gamma_l(\theta) + 
\gamma_s(\theta)}{\gamma_l(0) +\gamma_s(0)}
\end{equation}
where $\gamma_c$, $\gamma_l$ and $\gamma_s$ are the bistatic coefficients of
the coherent, diffuse and single scattering contributions respectively. 
$\theta$ is the angle of scattering and $L$ represents the set of
transport lengths in the medium. The set of relevant transport parameters
are the mean scattering length($l_s = 1/\langle \rho \sigma_{s} \rangle$)
defined as the reciprocal of the average of the product of the single
particle scattering cross-section ($\sigma_s$) and the density of
scatterers ($\rho $)(the probability of a photon remaining unscattered
after traversing a distance z in the medium is $\exp (-z/l_s)$), 
the inelastic length ($l_i$) defined as the
reciprocal of the absorption coefficient ($\alpha$) (transmitted intensity
$I_{t}(z) = I_{0} \exp (-z/l_{i})$ in the absence of scattering), the anisotropy
factor ($g= \langle \cos \theta \rangle$) which is defined as the average
of the cosine of the scattering angle for a single scattering event and
the mean transport length ($l_{t}$) which is a measure of the distance in which,
the direction of the photon's motion becomes uncorrelated with its initial
direction and is related as $l_{t} = l_{s}/(1 - g)$.
The bistatic coefficients for the diffuse and coherent intensities are
traditionally described by summing up the ladder diagrams and the most
crossed diagrams respectively in a systematic perturbation of the
intensity propagator \cite{mark}.  The bistatic coefficient for the single
scattering contribution for isotropic scatterers assuming normal incidence
is given by
\cite{ishimaru}
\begin{eqnarray}
\gamma_s(\theta) & = & \frac{4 \pi}{A} \int_{0}^{L} \exp\left[\frac{-x n_{0}(
\sigma_{s}+\sigma_{a})}{\cos\theta}\right] \left( \frac{n_{0}\sigma_{s}}{4\pi}
\right)
\exp\left[ -x n_{0}(\sigma_{s}+\sigma_{a}) \right] dx\cdot A  \nonumber \\
 & = & \frac{a \cos\theta}{ 1 + cos{\theta}}
\left\{ 1 - \exp\left[-b \left( 1 + \frac{1}{\cos{\theta}}
\right)\right] \right\}
\end{eqnarray}
where $A$ is the area of the target, $n_{0}$ is the density of scatterers, 
$\sigma_s$ is the single particle total scattering
cross-section, $\sigma_a$ is the absorption cross section,
$a$ is the albedo defined as the ratio of the total scattering 
cross-section to the extinction cross-section[$\sigma_{s}/(\sigma_{s} +\sigma_{a})$] 
and $b ( = n_{0}(\sigma_{s}+\sigma_{a})L)$ is the optical 
thickness of the slab.
However for the case of anisotropic scatterers, because it is the backscattering
coefficient which would contribute to the measurement, $a$ would be modified as
\begin{equation}
a = \frac {\sigma_b}{\sigma_s + \sigma_a}
\end{equation}
where $\sigma_b$ is the single particle back-scattering
coefficient.\\
 
When the single scattering contribution is supressed by using an
isolator \cite{etemad2} and the helicity preserving channel for
circularly polarized light is detected from time invariant
random media, the enhancement is exactly 2 in the weak disorder
regime \cite{wiersma2}. However if the detection is carried out
in the linear polarization preserving channel, the enhancement
factor will always be less than 2. This is due to the presence
of single scattered events which do not contribute to the 
enhancement in the back-scattering.
In earlier work \cite{yoo1,yoo2,yoon}, the contribution of 
single scattering was ignored.
Though Eddowes {\it et. al.} \cite{eddowes} suggest that the
experiment must be performed with circularly polarized light, it
has not been examined whether the single scattered contribution can
be used to estimate the anisotropy factor. The dependence of $a$
(normalized to unity for isotropic lossless scatterers) with the
anisotropy factor $g$ (obtained by varying the scatterer size), 
is shown in figure-1. 
The cross-sections
and the anisotropy factor were computed using Mie scattering
theory for spherical particles and a program based on the BHMIE
code given in \cite{bohren}. 
A value of 1.33 for the refractive index of the medium (water) and
1.41 for the refractive index of the particles (typical of
tissues\cite{wilson}) is used.  
Note the presence of resonances for monodisperse
particles in figure-1a. 
In figure-1b, the cross-sections and the $g$
factors were then averaged over a particle size dispersion of
30\% assuming a Gaussian distribution in order to mimic the
experimental situation where the particles are polydisperse . 
Slightly different values of the
refractive indices did not lead to much change, as the averaging
process washed all resonances present.  As one can notice, the
single scattering contribution decreases sharply with the
increase in anisotropy and becomes very small for large
anisotropy factors. For $g\raisebox{-0.6 ex}{\simgeq} 0.6$, 
the single scattering contribution 
from lossless scatterers is barely $2 \%$ of the total back-scattered 
intensity and would  hardly be measurable. This unfortunately implies 
that the anisotropy factor of biological tissues (typically 
in the range 0.8-0.95
\cite{wilson}) would hardly be measurable by this technique. \\  

The CBS peak shape is reasonably well described in the diffusion
approximation \cite{mark,akkermans}. In this work, the expressions  
derived in the diffusion approximation for an incident plane wave
on a slab of isotropic scatterers and with a finite thickness,
have been used \cite{mark} (See Appendix for the expressions). 
By replacing the mean scattering length ($l_{s}$) in these expressions by 
the mean transport length ($l_{t}$), these expressions have been 
shown to be valid for anisotropic scatterers as well\cite{akkermans2}. \\

It is essential to observe the effects of absorption on the CBS peak. 
The main effect of absorption is a rounding of the central cusp of the 
peak in an angular range $\theta_a \sim \lambda / \sqrt{l_{t} l_{i}}$.
This corresponds to the extinction of the longer path lengths greater than 
$\sqrt{l_t l_i}$ due to the absorption. The other effect of absorption is to
trivially reduce the enhancement factor because the single scattering 
events become relatively more important as absorption increases.
The CBS peak is also rounded by the 
convolution with the instrumental response function and unless the 
angular range $\theta_a$ of the rounding due to absorption is larger than
the instrumental resolution, the absorption would not be measurable 
accurately. For the case of $\theta_a = 0.18 mrad$ our experimental 
resolution, $\lambda = 0.6328 \mu m$ and $l_t = 670 \mu m$ a minimum value 
absorption coefficient $\alpha \sim 0.6 cm^{-1}$ can just be discerned. 
Though much smaller absorption
coefficients in principle can be measured by improving the angular 
resolution, even assuming ideal optics, the resolution eventually is
limited by diffraction due to the finite beam size. For our 
beam size of 5.5 mm, a beam
divergence ($\theta_d = \lambda / \pi R$ where R is the beam radius)
of about $75 \mu rad$ is present, yielding a minimum measurable $\alpha$ of 
about $0.1 cm^{-1}$ at a $l_t$ of $700 \mu m$. Trying to increase the beam 
size further is impractical due to problems of inhomogeneity of tissue    
samples. Hence 
accurate information would be possible only for samples with moderate 
absorption ($\alpha > 1 cm^{-1}$). However  
this is by no means a fundamental limit of the technique and depends 
on the transport 
length of the sample and the resolution of the experimental setup.\\

\section{Experimental setup and procedures}

The experimental setup used is schematically depicted in
figure-2.  A 5mW He-Ne laser at 632.8nm was expanded to a beam diameter of
5.5 mm and collimated to diffraction limit. The collimation was
checked by shear interferometry. Following the standard
practice, the CBS light was viewed through a non-polarizing
50-50 beam-splitter with a small wedge.  A computer controlled
CCD array (752 X 244 pixels, model EDC-1000HR, Electrim
corporation, U.S.A) was placed at the focal plane of a positive
lens with a focal length of 20 cm, to analyze the angular peak.
Each pixel on the CCD in this configuration corresponds to
0.06mrad.  The laser beam was linearly polarized by placing a
polarizer just before the beam-splitter. Another polarizer was
placed in front of the CCD to record images in the polarization
preserving channel. An aperture slightly larger than the input
beam was placed just behind the beam-splitter to avoid ghost
images from the AR coated surface of the beam-splitter. The free
beam going through the beam-splitter was carefully damped using
ND filters. \\

The setup was aligned by placing a mirror in place of the sample
and making the reflected beam go back into the laser. The beam
transmitted by the beamsplitter and focussed by the lens was
scanned by the CCD at the focal plane and the intensity profile
so obtained was used as the resolution curve characterizing the
system response. This was well described by a gaussian having a
FWHM of 0.17 mrad. The tissue phantom used was commercially
available boiled and skimmed milk. The concentration of the milk
was varied by adding distilled water. To prepare samples with
different absorptions, the milk was doped with known
concentrations of methylene blue dye. Methylene blue is a water
soluble dye having a broad absorption band in the red region
with two peaks at 610nm and 664nm The extinction coefficient of
the methylene blue solutions which were added to the milk, was
measured by a Shimadzu spectrophotometer.  The dye-doped
colloids thus had a well determined inelastic length.  The milk
did not change its scattering properties measurably in a time of
about 3 hours. This was confirmed by recording the CBS profiles
at different times and comparing them. All the experiments were
hence carried out on the same milk sample within this time
frame. The tissue phantom was taken in a cuvette of 10mm path
length. It was placed slightly tilted to the incident beam so
that the specular reflection was well away from the
backscattered direction.  The Brownian motion of the milk
particles caused an ensemble average over the speckle and sharp
symmetric peaks were observed. CBS intensities for samples of
different scatterer concentration and absorption were
recorded.\\

The CBS peaks from milk suspensions were very sharp (FWHM $\sim 1mrad$) and
consequently affected by the finite instrumental response. Hence
in order to fit the parameters, the theoretical profiles were
first convolved with the system resolution curve using FFT
routines. The resulting curve was then least squares fitted to
the experimental points treating $l_t, l_i$ and $\gamma_s$ as parameters
to be fitted using standard NAG library routines.  Deconvolution
of the experimental points was avoided as this was found to be a
noisy process. \\

To supplement the CBS measurements, independent measurements of
the transport parameters were made by measuring the transmission
of light through milk solution as a function of the optical
thickness or the concentration of milk. The milk was taken in a
cuvette of 5mm path length.  A 2mW He-Ne laser was used as the
source.  The light exiting at the back of the cuvette was imaged
onto a photodiode, whose signal was detected by a lockin
amplifier controlled by a computer. A separate photodiode was
used to simultaneously record the laser intensity
fluctuations. We have also undertaken angle-resolved scattering
measurements for estimating the value of the anisotropy factor
at the same wavelength. The sample was taken in a 1.0 mm cylinderical 
cuvette and is diluted enough, so as to get into the single
scattering regime. The transmission of the sample was coupled to
the PMT through an optical fibre kept on a rotating stage such
that the distance from the sample to the fibre tip does not change
and
is read by a digital storage oscilloscope. A reference detector
monitored the laser intensity fluctuations. 
Measurements were avoided in the lower angle region upto 8
degrees.

\section{Results}

Figure-3 shows the CBS profiles obtained from two different
concentrations of milk. The solid lines shown are the best fits
to the corresponding experimental points. As expected the CBS
profiles became narrower with increasing transport length.  The
enhancement factor for the sample with longer transport length
is considerably reduced due to the effect of convolution with
the system response. The fits are quite good with a figure of
merit ($\chi^{2}/N$ where N is the number of degrees of freedom
for the fit) of about 1.2. The variation of the inverse of the
transport length with the concentration is shown in the inset of
figure-3. The $l_t^{-1}$ increases linearly with concentration
as expected. The transport length is well defined and the
numerical routines used quickly converge to the correct value to
an accuracy of $\pm 5\%$ in a few iterations. However for the
plain milk solutions any absorption coefficient less than
1.5$cm^{-1}$ could be fitted with marginal changes in the
transport length. This is because the intensity measurement
accuracy is limited to about $3\%$ by the CCD noise and the
changes in the profile caused by small amounts of absorption are
of the same level. \\

Figure-4 shows the CBS intensity obtained from
dye-doped milk with a pre-determined absorption coefficient of 5
$cm^{-1}$. The fitted curve is for an absorption coefficient of 
$(5.2\pm 0.2)cm^{-1}$ and the elastic transport lengths are within
reasonable variation for the pure milk ($l_t = 625 \pm 30 \mu m $) and
the absorbing sample ($l_t = 670\pm 30\mu m$). At higher absorption, the entire 
peak appears broadened because the angular range of the rounded cusp is 
$\theta_a \sim 0.6 mrad$ while the peak width itself is of the same order. 
Also the height of the peak seems to be relatively unaffected by the 
absorption even though the weight of single scattering is enhanced. 
This counter-intuitive behaviour occurs because of the decreased
effect of the instrumental response on the broader peak compared to that 
of the lower absorption narrow peak. This is clearly brought out in figure-5 
where the theoretical shape (assuming same $\gamma_s$ as from experiment) 
for an infinite resolution is plotted. The high absorption peak clearly 
shows a lower enhancement. Thus the finite resolution and the relative 
weight of single scattering act in opposite directions on the enhancement 
factor with increasing absorption.
The inset in Figure-4
shows the CBS profile and the theoretical fit from a sample with
higher absorption ($\alpha = 10 cm^{-1} $). The backscattered
intensity level was small necessitating large exposure times,
increasing the CCD noise.  However the theoretical fit again
correctly yielded the absorption coefficient within an accuracy
of $\pm10\%$. We studied the CBS in dye doped milk having
absorption coefficients upto 30$cm^{- 1}$. However, the noise
at these absorption levels was high (around 30 $\%$ of the peak
intensity) and would appear to limit the use of this technique to
measure larger absorption coefficients. Anyhow at higher levels 
of absorption, the expressions for the CBS peak shape 
derived within the diffusion approximation 
will no longer be valid. \\ 

Using the best fitted values of $\gamma_s$, the enhancement 
factors of the deconvolved curves  if the plain milk-water 
solutions (no dye) are found to be in the range 
of 1.82 to 1.86. This reduction in the enhancement factor 
is lesser than the case of isotropic scatterers in a finite slab
of same thickness with the same $l_t$ and $l_i$ for which the 
theoretical enhancement factors are between 1.76 to 1.80. These 
values are in agreement with those reported by Wolf {\it et. al.}\cite{wolf2},
but as observed by them the reduction in the enhancement
factor appears much too large for such forward scattering media.
Other processes such as recurrent multiple scattering \cite{wiersma2}
which reduce the enhancement are not important at such large mean free
paths present here. 
The lowering of the enhancement factor could also be caused by
the finite beam effects and the gaussian intensity profile 
of the laser beam. However we do not
believe this to be the reason as the beam size is reasonably large 
(5.5 mm FWHM $\sim 10 l_{t}$). Using the equation(60) of Jakeman\cite{jakeman} for 
Gaussian beams, we estimate that the reduction in the enhancement factor 
can only be in the range of 0.01 to 0.04 for the different $l_t$ used. 
In spite of our best efforts, higher enhancements 
were not observed. The CBS peak from a piece of white paper with a broad peak (FWHM 
of $10 mRad$) was observed to have an enhancement of  only about 1.85 without 
deconvolution. The exact reason for this lower enhancement is not clear, but 
it could be due to some of the laser light being scattered by the non-ideal
optical elements, cuvette walls, aggregates and dust which cause an 
additional component to the diffuse background. This would, however, be 
sample-independent. Assuming this additional contribution to be uniformly distributed, 
we have attempted to quantify it by studying the enhancement factor with
respect to absorption in the sample. We conclude from this procedure that 
the ratio $\sigma_{b}/\sigma_{s} \raisebox{-0.6 ex}{\simleq}0.1$ and that the g-factor of milk is 
greater than 0.55, which is what we expect from our theoretical considerations
given earlier. \\

The transport properties can also be obtained from the plot of
transmission versus the optical thickness \cite{ishimaru}. The
results of the transmission measurements through milk-water
suspension is shown in figure-6. At low densities (region A),
the transmission falls exponentially depending on the extinction
coefficient, the slope of the logarithmic transmittance being
given by $\sigma_s +\sigma_a$. In the multiple scattering domain
(region B), the transmission falls inversely with concentration
or optical thickness. At large optical thickness or
concentration (region C), the transmission falls exponentially
depending on the absorption coefficient. The slope of the
logarithmic transmittance is $\sigma_a$ in this region. These
yield a value of the scattering length ($l_s$) 
of 130$\mu m$
and an absorption coefficient of $1.6cm^{-1}$ for milk at the
largest concentration used. As it can be noticed the absorption
coefficient lies just in the region where the CBS profiles
become insensitive to absorption. 
Using a value of $l_t = 470 \mu m$
obtained from the CBS measurements for this concentration and
$l_s = 130 \mu m$ from the transmission experiment, a value of
$g=0.74$ is obtained which appears reasonable for such forward
scattering media. The inset in figure-6 shows the  
angle-resolved scattering data, together with a theoretical fit
to the Henyey-Greenstein function (solid line) with $g$ as the
free parameter to be fitted. The fit yielded a
value of about 0.70 for the anisotropy factor ($g $) with an error of
about 5\%.\\ 

\section{Conclusions}

In conclusion, we have investigated the application of Coherent
backscattering to measure the light transport parameters in
tissues using tissue phantoms (milk and dye doped milk), paying
particular attention to the limitations of the technique. This
method yields very good estimates of the transport length but it
cannot be used to estimate the inelastic length when the
absorption of the sample is small ($\alpha < 1.0 cm^{-1}$) for
the long transport length typical of tissues. The maximum
measurable absorption coefficient ($ \alpha \sim 30cm^{-1} $)
with our setup appears to be limited by the sensitivity and
noise levels of the detector used.  
It has been pointed out that in principle this technique
can also be used to measure the anisotropy factor($g$) by
estimating the single scattering contribution to the angle
independent diffuse intensity. The single scattering contribution 
becomes very small for $g \raisebox{-0.6 ex}{\simgeq} 0.6$ ($\sim 2\%$  of the total 
backscattered intensity) and the technique becomes insensitive 
for larger values of $g$. Thus it does not seem to be a suitable
technique to estimate the $g$-factor of biological media
which typically are highly anisotropic ($g \sim 0.8$ to $0.95$).
Using the value of the
transport length obtained from CBS measurements and scattering
length from transmission measurements, a value of 0.74 has been
obtained for the anisotropy parameter of milk. This value is
confirmed independently by angle-resolved scattering measurement 
where a value of 0.70 is obtained.\\

\section*{Appendix}
For the convenience of the reader, the result of Ref.\cite{mark}
for the angular shape of the CBS peak from a slab of finite thickness 
is reproduced here. \\

Let us define $\beta = \sqrt{ \left( l_{t} l_{i}/3 + q_{\perp}^{2} \right)}$, 
$q_{\perp} = (2 \pi/\lambda) \sin\theta$, $\delta = (2\pi/\lambda) (1-
\cos\theta )$, $ \eta = (1+1/\cos \theta)/ 2l_{t}$ and $z_{0} = 0.71 l_{t}$,
where $\theta$ is the backscattering angle, $\lambda $ is the 
wavelength of the light, $l_{t}$ is the mean transport
length and $l_{i}$ is the inelastic length. Now the angular shape of 
the CBS peak in our notation is given by
\begin{eqnarray}
\gamma_{c} (\theta, l_{t}, l_{i}) & =& \frac{3 e^{-\eta L}}{2 l_{t}^{3} \beta
\sinh [\beta (L+2z_{0})]} \frac{1}{(\eta^{2} + \delta^{2} - \beta^{2} )^{2} 
+ (2 \beta \delta)^{2}}  \nonumber\\
&\{ & \frac{2\beta}{\eta} 
(\beta^{2} + \delta^{2} - \eta^{2} ) \sinh [\beta (L+2z_{0})] \sinh (\eta L)
+ 2 (\eta^{2} + \delta^{2} + \beta^{2} ) \cos (\delta L)   \nonumber\\
&+& 2 
(\eta^{2} + \delta^{2} - \beta^{2} ) \cosh [\beta (L+2z_{0})] \cosh (\eta L)
+ 4 \beta \eta \sinh (\beta L) \sinh (\eta L)  \nonumber\\
&-& 2 (\eta^{2} + \delta^{2} 
+ \beta^{2} ) \cosh (\beta L) \cosh (\eta L) - 2 (\eta^{2} + \delta^{2} 
- \beta^{2} ) \cosh (2\eta z_{0}) \cos  (\delta L)  \nonumber\\
&-&  \beta \eta 
\sinh (2 \beta z_{0}) \sin (\delta L) ~~\} ,
\end{eqnarray}
where L is the thickness of the slab. It should be noted that the
diffuse part of the backscattered light only has a weak kinematic 
dependence on the angle
and the $\gamma_{l}(\theta,L) \simeq \gamma_{c}(0,L)$. In the 
exact backscattering direction $\gamma_{l}(0,L) = \gamma_{c}(0,L)$.\\

\section*{Acknowledgments}
One of us (SAR) would like to sincerely acknowledge Diederick
Wiersma (European Laboratory for Non-Linear Spectroscopy,
Florence, Italy) for encouragement and for patiently answering
all our questions over the e-mail. Both the authors thank
P.K.Gupta for support and encouragement, and  N.Ghosh for his
help with the angle-resolved scattering measurements. \\

\newpage
\section*{Figure Captions}
\vspace{0.5cm}

\noindent {\bf Figure-1} : Plot of the ratio of the backscattering 
cross-section($\sigma_b$) to the total scattering
cross-section($\sigma_s$) as a function of the anisotropy factor
($g$)(normalized to unity for isotropic scattering.  These
quantities were calculated using Mie scattering theory for
homogeneous spheres. Figure-1a is for monodisperse particles and 
figure-1b shows the dependence averaged over a particle size 
distribution of 30\%.   \\

\noindent {\bf Figure-2} : Experimental set-up for observing the 
coherent backscattered light. L1-L3 : lenses; M : mirror; A1,A2
: apertures; P1, P2 : polarizers; BS : beam-splitter; BD :
beam-dump; S : sample.  \\

\noindent {\bf Figure-3} : Measured CBS intensities for milk of two 
different concentrations. The solid lines are the calculated fits to the 
corresponding experimental points. The fitted profiles are for : 
i) $l_{t}$ of $(490 \pm 20)\mu m (\bullet)$ and $ (1100 \pm 50)\mu m 
(\diamond)~~$  
ii) $l_i$ of $(15000 \pm 5000)\mu m(\bullet)$ and 
$ (13000 \pm 5000)\mu m(\diamond)~~$ 
iii) $\gamma_{s}$ of $(0.3 \pm 0.02)$ for both fits. The inset 
shows the variation of inverse of transport length with the concentration.\\

\noindent {\bf Figure-4} : CBS profiles from methylene blue doped milk 
for different absorption coefficients. Solid lines are the calculated best 
fits to the corresponding experimental points with 
i) $l_{t}$ of $(670 \pm 30) \mu m (\bigtriangleup )$ and 
$(680 \pm 30)\mu m (\bullet )$, ii) $l_{i} > 7000 \mu m (\bigtriangleup)$
and $l_{i}$ of $(1900 \pm 200) \mu m (\bullet)$ respectively. The 
inset shows the CBS profile from dye doped milk at higher dye concentration 
($\alpha \sim 10 cm^{-1}$ and $l_t  = (630 \pm 50) \mu m$).\\ 

\noindent {\bf Figure-5} : The CBS intensities of the best-fits 
in figure-4 for an infinite resolution.  The enhanced weight of the 
single scattering contribution for the sample with higher absorption 
resulting in lower enhancement is clearly seen.\\

\noindent {\bf Figure-6} : Variation of the transmittance of milk as a 
function of the effective optical thickness of milk-water
solution. The exponential fits in regions A and C are shown. The
slopes of the logarithmic transmittances are 8.01 and 0.16 and
yield values of 130$\mu m$ for the scattering length and
$1.6cm^{-1}$ for the absorption coefficient respectively, at the
largest concentration of the milk. Inset shows the angle resolved scattering data.
The solid line is a theoretical fit to the Henyey-Greenstein function giving a 
value of $0.7 \pm 0.04$ for $g$.
\\

\begin{references}
\bibitem{error} These authors have described the total backscattered
intensity in terms of the sum of a coherent intensity arising
from an absorbing medium and a diffuse intensity from a non-absorbing medium.
As the intensity reduces
with absorption, the coherent contribution becomes lesser
reducing the enhancement factor in their expression. The single scattering
contribution was not included either.
\end{references}
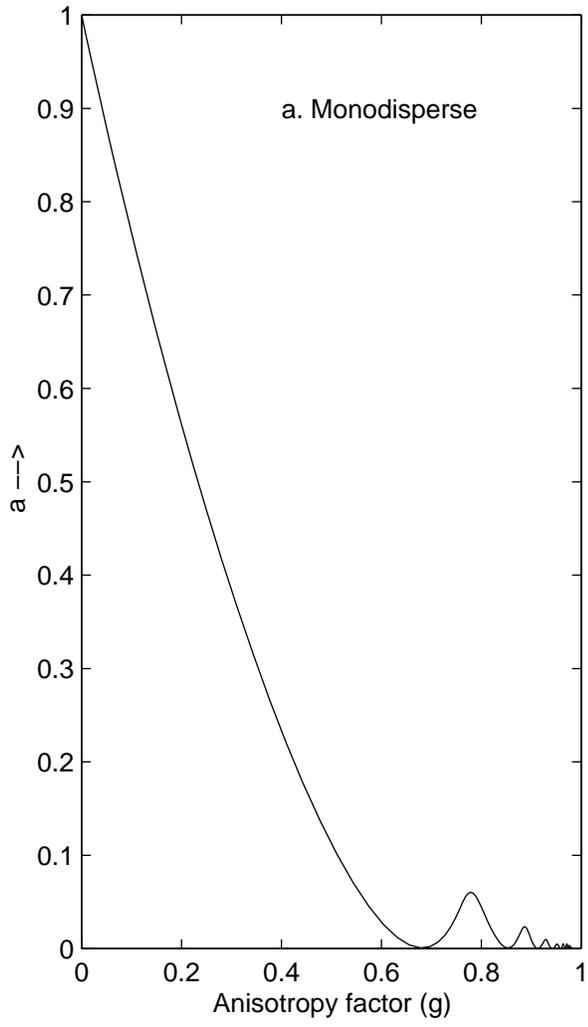
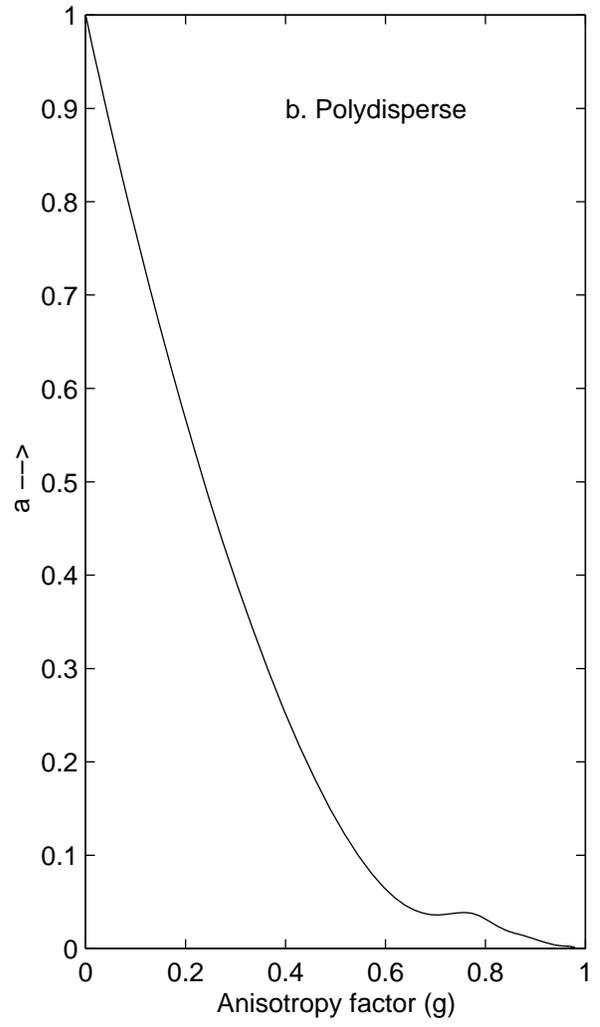
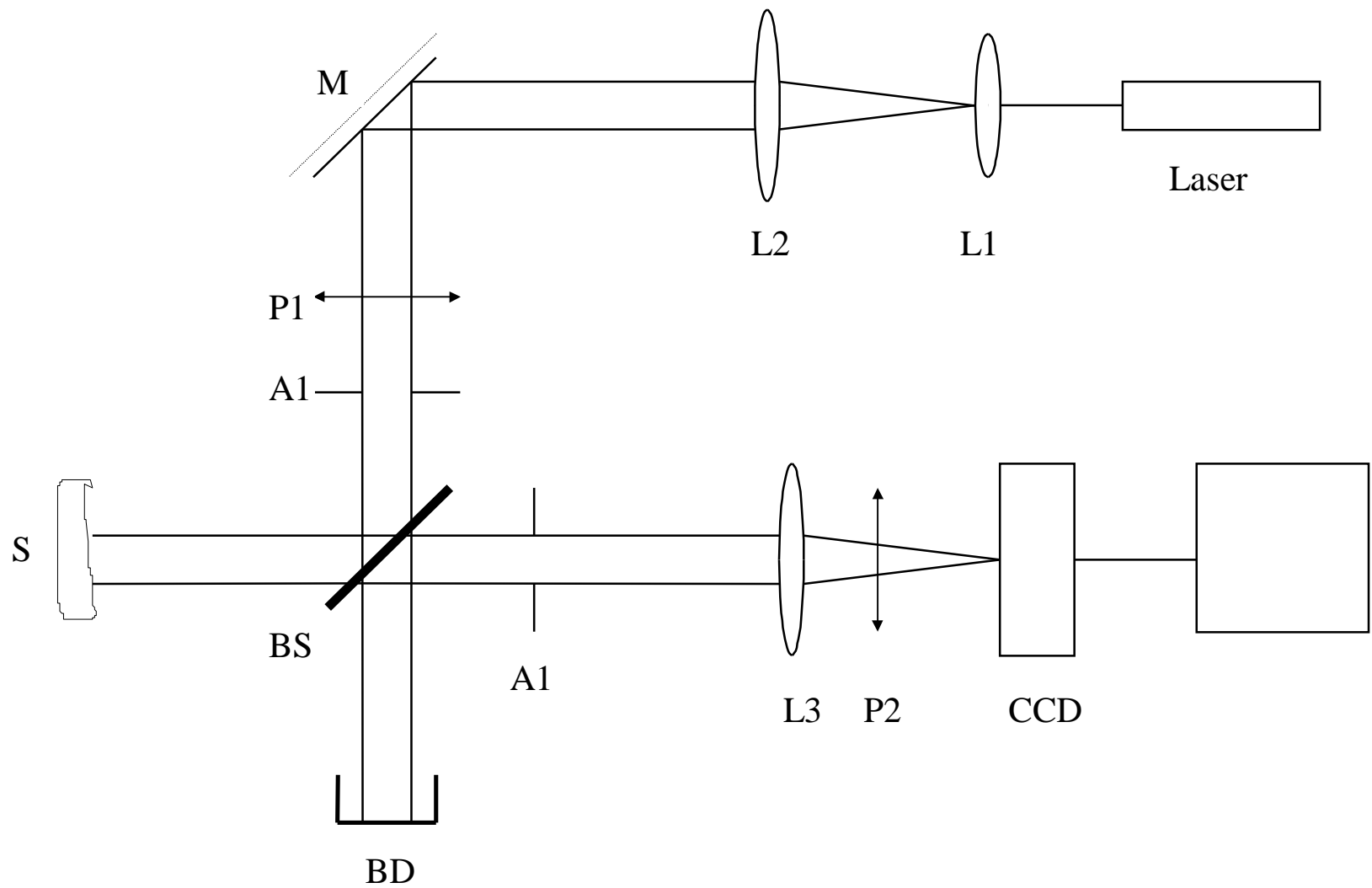
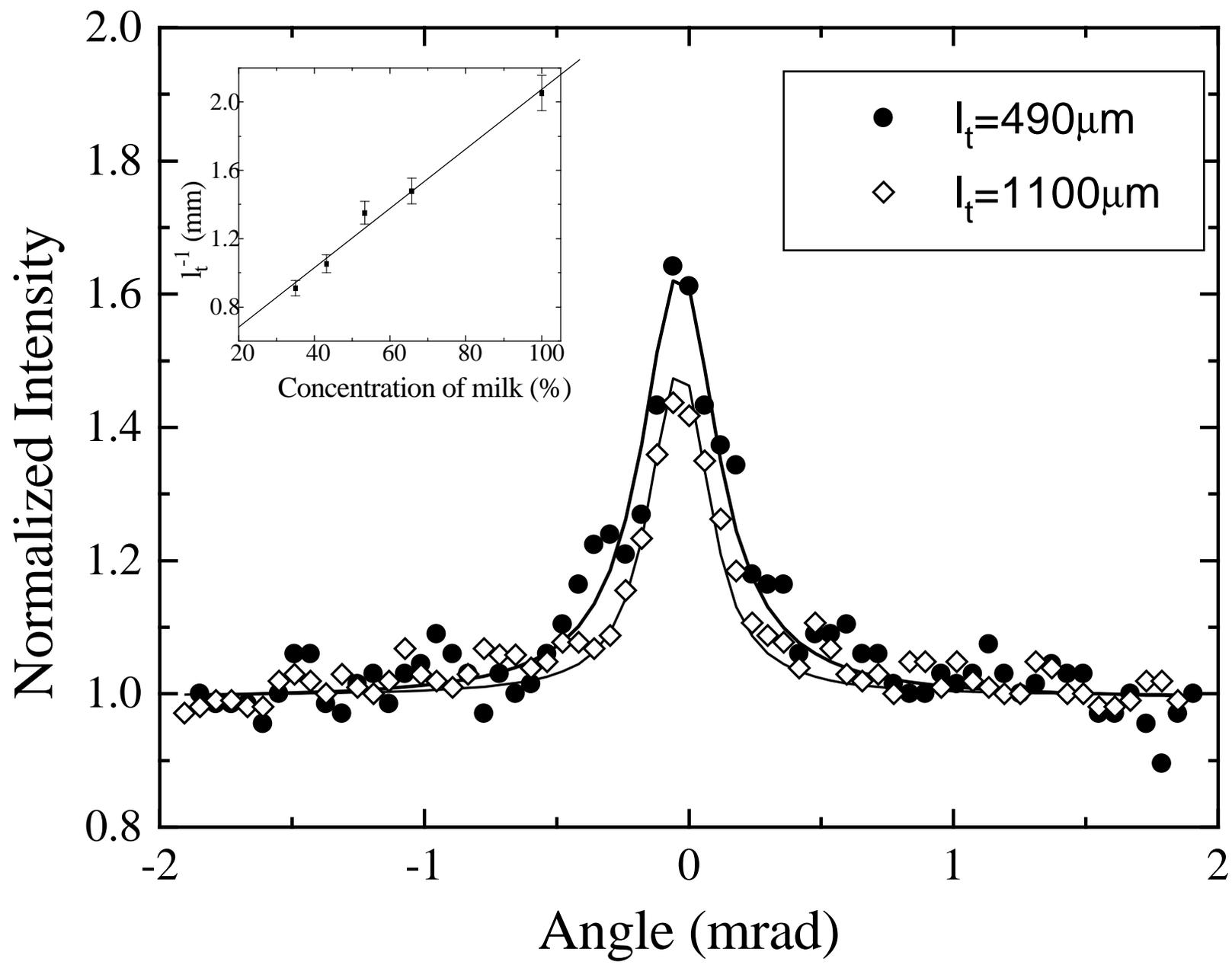
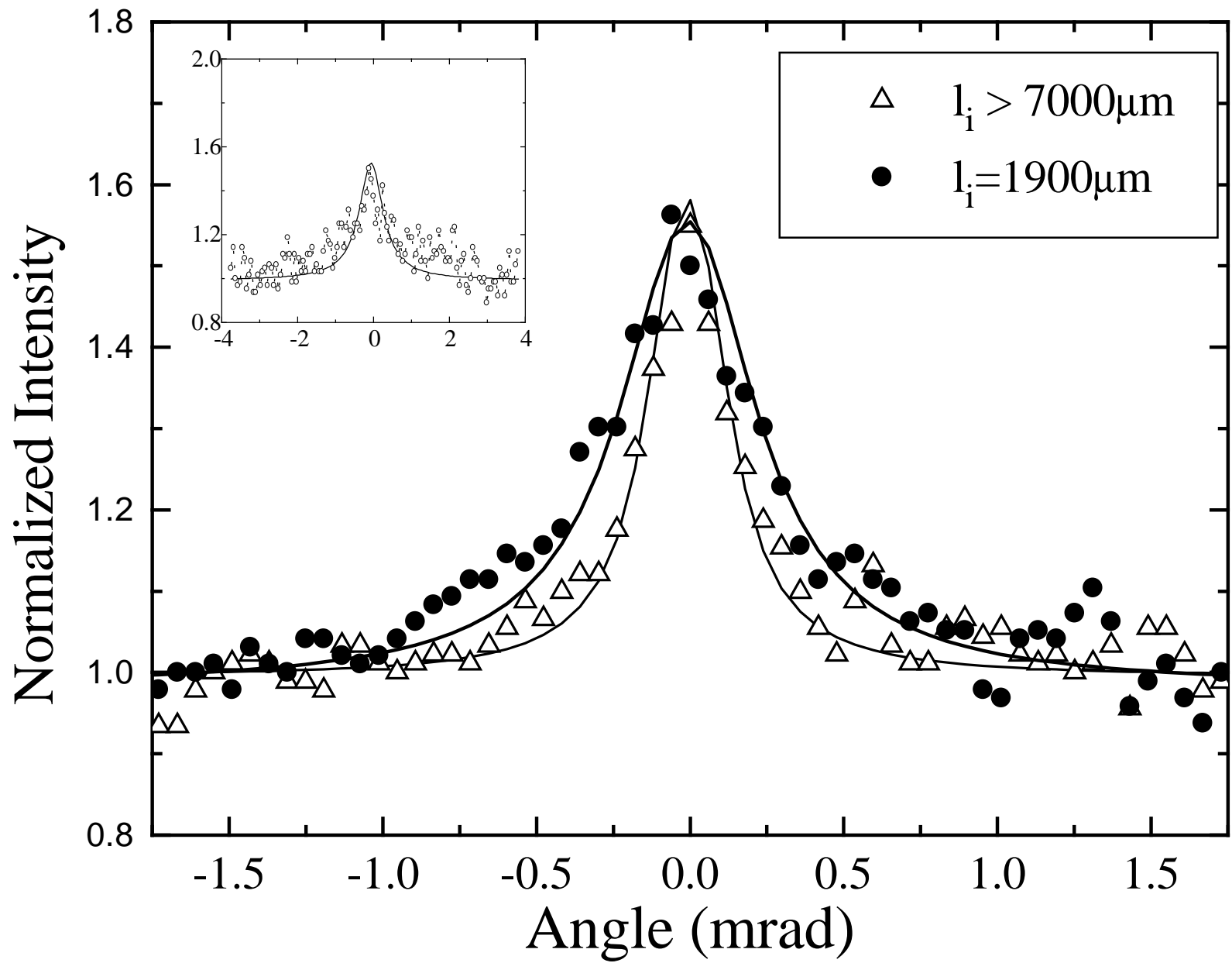
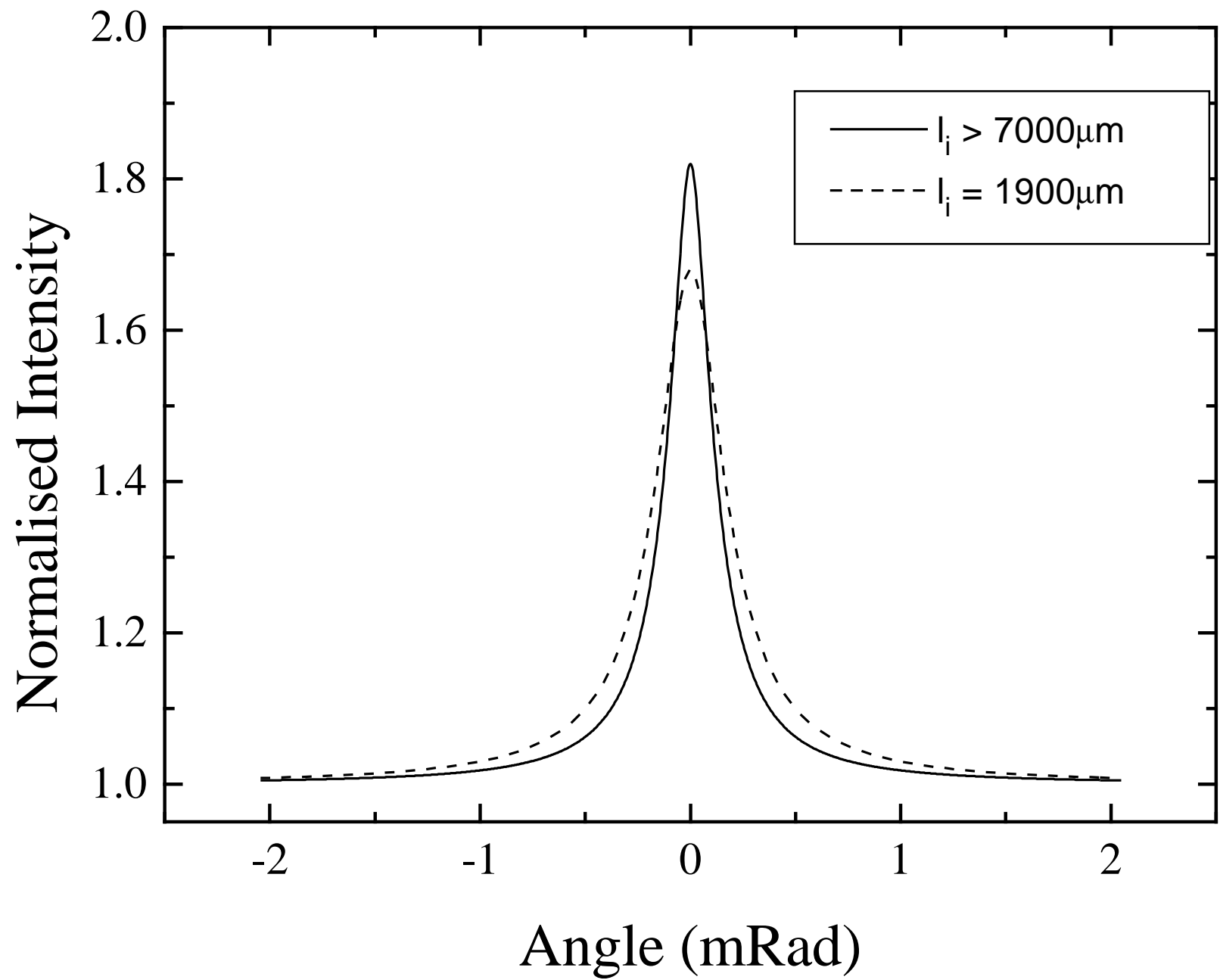
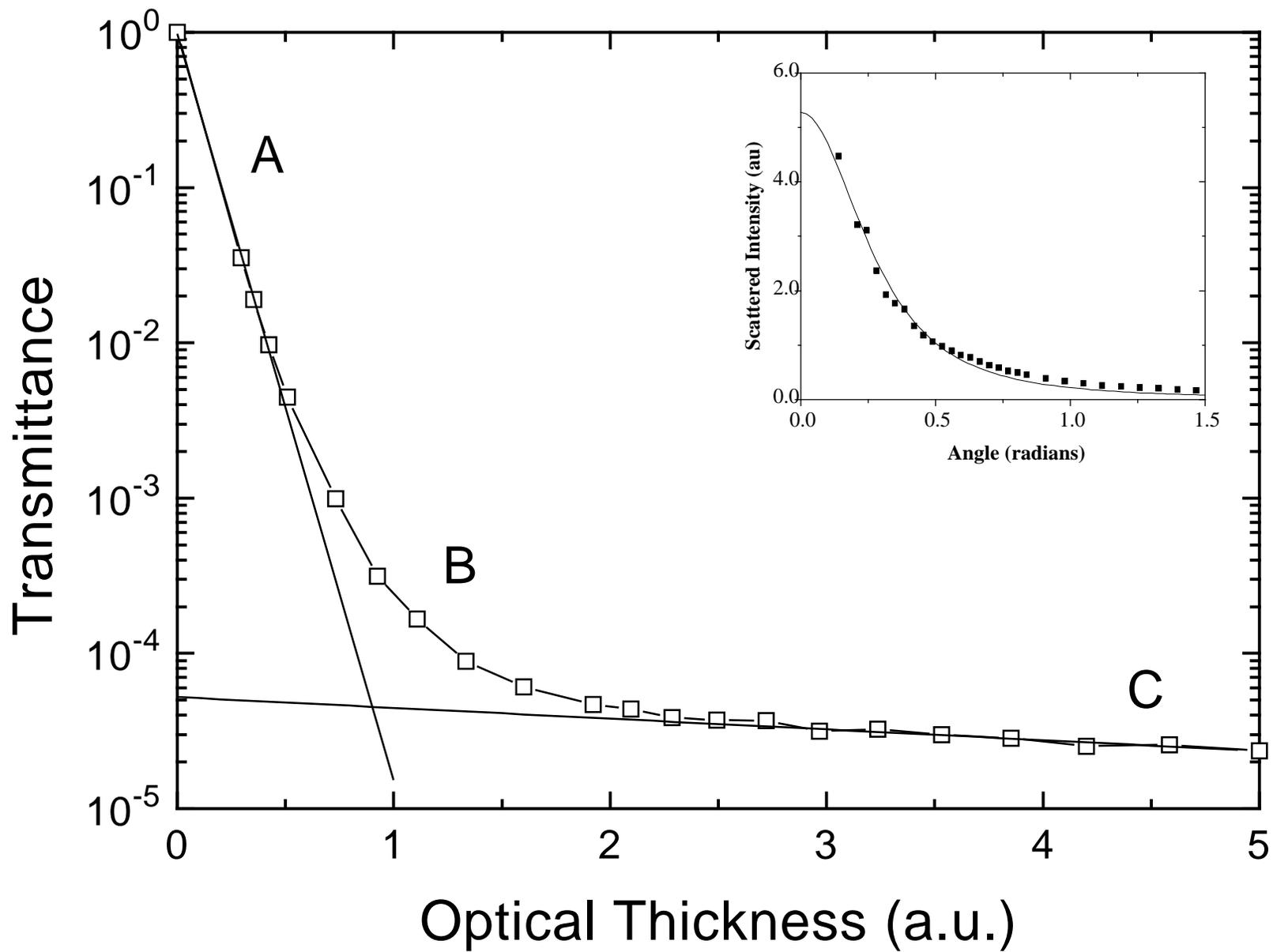
\end{document}